\documentclass[letterpaper,twocolumn,10pt]{article}
\usepackage{usenix-2020-09}
\usepackage{inconsolata}
\usepackage{amsmath}
\usepackage{graphicx}
\usepackage{subcaption}
\usepackage{enumitem}
\usepackage{makecell,multirow}
\usepackage{titlesec}
\usepackage{nopageno} 
\usepackage{titling}

\usepackage[backref=false,doi=false,eprint=false,isbn=false,url=false,
  maxbibnames=5,style=numeric,sortcites]{biblatex}
\AtEveryBibitem{
 \clearlist{address}
 \clearfield{eprint}\clearfield{isbn}\clearfield{issn}
 \clearlist{location}\clearfield{month}\clearfield{series}
 \ifentrytype{book}{}{\clearlist{publisher}\clearname{editor}}
}
\hypersetup{colorlinks,linkcolor={teal!80!black},citecolor={purple!70!black}}
\setlist[itemize]{leftmargin=16pt,topsep=2pt,itemsep=2pt,partopsep=0pt,parsep=0pt}
\titlespacing*{\paragraph}{0pt}{2pt plus 6pt minus 0pt}{1em plus .5em}

\begin{document}
\title{\Large \bf REMIX: Efficient Range Query for LSM-trees}
\author{\rm
{\rm Wenshao Zhong$^\star$\quad Chen Chen$^\star$\quad Xingbo Wu$^\star$\quad Song Jiang$^\dagger$}\\
{\normalsize $^\star$University of Illinois at Chicago\quad
$^\dagger$University of Texas at Arlington}}

\date{}
\maketitle

\begin{abstract}
LSM-tree based key-value (KV) stores organize data in a multi-level structure for high-speed writes.
Range queries on traditional LSM-trees must seek and sort-merge data from multiple table files on the fly,
which is expensive and often leads to mediocre read performance.
To improve range query efficiency on LSM-trees,
we introduce a space-efficient KV index data structure, named REMIX,
that records a globally sorted view of KV data spanning multiple table files.
A range query on multiple REMIX-indexed data files can quickly locate the target key using a binary search,
and retrieve subsequent keys in sorted order without key comparisons.
We build RemixDB, an LSM-tree based KV-store that adopts a write-efficient compaction strategy
and employs REMIXes for fast point and range queries.
Experimental results show that REMIXes can substantially improve range query performance
in a write-optimized LSM-tree based KV-store.

\end{abstract}
\section{Introduction}

Key-value stores (KV-stores) are the backbone of many cloud and datacenter services,
including social media~\cite{linkbench13,fb-workload12,fb-workload20},
real-time analytics~\cite{rta15,olxp15,realtime16}, e-commerce~\cite{dynamo07},
and cryptocurrency~\cite{mLSM18}.
The log-structured merge-tree (LSM-tree)~\cite{lsmtree96} is the core data structure of many KV-stores~\cite{rocksdb,leveldb,bigtable08,dynamo07,cassandra,scylladb}.
In contrast to traditional storage structures (e.g., B+-tree) that require in-place updates on disk,
LSM-trees follow an out-of-place update scheme which enables high-speed sequential write I/O.
They buffer updates in memory and periodically flush them to persistent storage to generate immutable table files.
However, this comes with penalties on search efficiency as keys in a range may reside in different tables,
potentially slowing down queries because of high computation and I/O costs.
The LSM-tree based designs represent a trade-off between update cost and search cost~\cite{wackybush19},
maintaining a lower update cost but a much higher search cost compared with a B+-tree.

Much effort has been made to improve query performance.
To speed up point queries, every table is usually associated with memory-resident Bloom filters~\cite{bloom}
so that a query can skip the tables that do not contain the target key.
However, Bloom filters cannot handle range queries.
Range filters such as SuRF~\cite{surf18} and Rosetta~\cite{rosetta20} were proposed
to accelerate range queries by filtering out tables not containing any keys in the requested range.
However, when the keys in the requested range reside in most of the candidate tables,
the filtering approach can hardly improve query performance,
especially for large range queries.
Furthermore, the computation cost of accessing filters can lead to mediocre performance
when queries can be answered by cache, which is often the case in real-world workloads~\cite{ycsb2010,fb-workload12,fb-workload20}.


To bound the number of tables that a search request has to access,
LSM-trees keep a background compaction thread to constantly sort-merge tables.
The table selection is determined by a compaction strategy.
The \emph{leveled} compaction strategy has been adopted by a number of KV-stores, including LevelDB~\cite{leveldb} and RocksDB~\cite{rocksdb}.
Leveled compaction sort-merges smaller sorted runs into larger ones to keep the number of overlapping tables under a threshold.
In practice, leveled compaction provides the best read efficiency but has a high write amplification (WA) due to its aggressive sort-merging policy.
Alternatively, the \emph{tiered} compaction strategy waits for multiple sorted runs of a similar size and merges them into a larger run.
Tiered compaction provides lower WA and higher update throughput.
It has been adopted by many KV-stores, such as Cassandra~\cite{cassandra} and ScyllaDB~\cite{scylladb}.
Since tiered compaction cannot effectively limit the number of overlapping tables, it leads to much higher search cost compared with leveled compaction.
Other compaction strategies can better balance the read and write efficiency~\cite{dosto18,wackybush19},
but none of them can achieve the best read and write efficiency at the same time.

The problem lies in the fact that, to limit the number of sorted runs,
a store has to sort-merge and rewrite existing data.
Today's storage technologies have shown much improved random access efficiency.
For example, random reads on commodity Flash SSDs can exceed 50\% of sequential read throughput.
New technologies such as 3D-XPoint (e.g., Intel's Optane SSD) offer near-equal performance for random and sequential I/O~\cite{optanecontract}.
As a result, KV-pairs do not have to be \emph{physically} sorted for fast access.
Instead, a KV-store could keep its data \emph{logically} sorted for
efficient point and range queries while avoiding excessive rewrites.

To this end, we design REMIX, short for \textbf{R}ange-query-\textbf{E}fficient \textbf{M}ulti-table \textbf{I}nde\textbf{X}.
Unlike existing solutions to improve range queries that struggle between physically rewriting data and performing expensive sort-merging on the fly,
a REMIX employs a space-efficient data structure to record a globally sorted view of
KV data spanning multiple table files.
With REMIXes, an LSM-tree based KV-store can take advantage of
a write-efficient compaction strategy without sacrificing search performance.

We build RemixDB, a REMIX-indexed LSM-tree based KV-store.
Integrated with the write-efficient tiered compaction strategy and a partitioned LSM-tree layout,
RemixDB achieves low WA and fast searches at the same time.
Experimental results show that REMIXes can effectively improve range query performance
when searching on multiple overlapping tables.
Performance evaluation demonstrates that RemixDB outperforms the state-of-the-art
LSM-tree based KV-stores on both read and write operations simultaneously.
\section{Background}

The LSM-tree is designed for high write efficiency on persistent storage devices.
It achieves high-speed writes by buffering all updates in an in-memory structure, called a MemTable.
When the MemTable fills up, the buffered keys will be sorted and
flushed to persistent storage as a sorted run by a process called \emph{minor compaction}.
Minor compaction is write-efficient because updates are written sequentially in batches
without merging with existing data in the store.
Since the sorted runs may have overlapping key ranges,
a point query has to check all the possible runs, leading to a high search cost.
To limit the number of overlapping runs, an LSM-tree uses a \emph{major compaction} process
to sort-merge several overlapping runs into fewer ones.

\begin{figure}[t]
    \centering
    \includegraphics[width=.85\columnwidth]{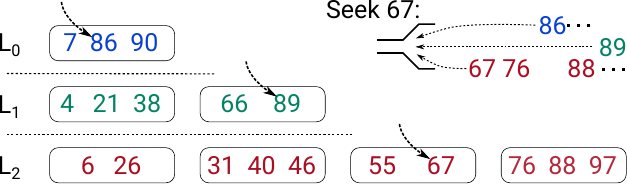}
    \caption{An LSM-tree using leveled compaction}
    \label{fig:leveled}
\end{figure}

A compaction strategy determines how tables are selected for major compaction.
The two most commonly used strategies are leveled compaction and tiered compaction.
A store using leveled compaction has a multi-level structure where
each level maintains a sorted run consisting of one or more tables.
The capacity of a level ($L_{n}$) is a multiple (usually 10~\cite{rocksdb}) of the previous one ($L_{n-1}$),
which allows a huge KV-store to be organized within a few levels (usually 5 to 7).
Leveled compaction makes reads relatively efficient, but it leads to inferior write efficiency.
Leveled compaction selects overlapping tables from adjacent levels ($L_{n}$ and $L_{n+1}$)
for sort-merging and generates new tables in the larger level ($L_{n+1}$).
Because of the exponentially increasing capacity, 
a table's key range often overlaps several tables in the next level.
As a result, the majority of the writes are for rewriting existing data in $L_{n+1}$,
leading to high WA ratios\footnote{
WA ratio refers to write amplification ratio,
or ratio of the amount of actual data written on the disk to the amount of user-requested data written.}
of up to 40 in practice~\cite{pebbles17}.
Figure~\ref{fig:leveled} shows an example of leveled compaction where each table contains two or three keys.
If the first table in $L_1$ (containing keys $(4,~21,~38)$) is selected for sort-merging with
the first two tables in $L_2$ ($(6,26)$ and $(31,40,46)$), five keys in $L_2$ will be rewritten.

\begin{figure}[t]
    \centering
    \includegraphics[width=.85\columnwidth]{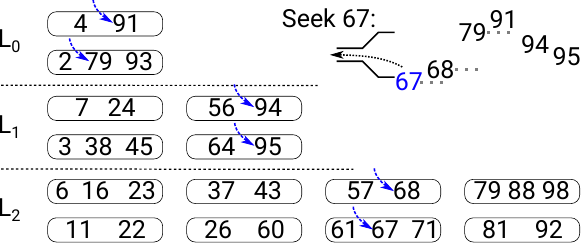}
    \caption{An LSM-tree using tiered compaction}
    \label{fig:tiered}
\end{figure}

With tiered compaction, multiple overlapping sorted runs can be buffered in a level, as shown in Figure~\ref{fig:tiered}.
The number of runs in a level is bounded by a threshold denoted by $T$, where $T>1$.
When the number of sorted runs in a level ($L_{n}$) reaches the threshold,
all sorted runs in $L_{n}$ will be sort-merged into a new sorted run in the next level ($L_{n+1}$),
without rewriting any existing data in $L_{n+1}$.
Accordingly, an LSM-tree's WA ratio is $O(L)$ using tiered compaction~\cite{monkey17}, where $L$ is the number of levels.
With a relatively large $T$, tiered compaction provides much lower WA than leveled compaction does with a similar $L$.
However, since there can be multiple overlapping sorted runs in each level,
a point query will need to check up to $T\times L$ tables, leading to a much slower search.

Range query in LevelDB/RocksDB is realized by using an \emph{iterator} structure
to navigate across multiple tables as if all the keys are in one sorted run.
A range query first initializes an iterator using a \emph{seek} operation with a \emph{seek key},
the lower boundary of the target key range.
The seek operation positions the iterator so that it points to the smallest key in the store
that is equal to or greater than the seek key (in lexical order for string keys), 
which is denoted as the \emph{target key} of the range query.
The \textit{next} operation advances the iterator such that it points to the next key in the sorted order.
A sequence of next operations can be used to retrieve the subsequent keys in the target range
until a certain condition is met (e.g., number of keys or end of a range).
Since the sorted runs are generated chronologically, a target key can reside in any of the runs.
Accordingly, an iterator must keep track of all the sorted runs.

Figure~\ref{fig:leveled} shows an example of seek on an LSM-tree using leveled compaction.
To seek key 67, a binary search is used on each run to identify the smallest key satisfying $key\ge seek\_key$.
Each identified key is marked by a cursor.
Then these keys are sort-merged using a min-heap structure~\cite{minheap}, and thus the key 67 in $L_2$ is selected.
Subsequently, each next operation will compare the keys under the cursors,
return the smallest one, and advance the corresponding cursor.
This process presents a globally sorted view of the keys,
as shown in the upper right corner of Figure~\ref{fig:leveled}.
In this example, all three levels must be accessed for the sort-merging.
Figure~\ref{fig:tiered} shows a similar example with tiered compaction.
Having six overlapping sorted runs, a seek operation is more expensive than the previous example.
In practice, the threshold $T$ in tiered compaction is often set to a small value,
such as $T=4$ in ScyllaDB~\cite{scylladb}, to avoid having too many overlapping sorted runs in a store.
\section{REMIX}

A range query operation on multiple sorted runs constructs a \emph{sorted view} of the underlying tables on the fly
so that the keys can be retrieved in sorted order.
In fact, a sorted view inherits the immutability of the table files and remains valid until any of the tables are deleted or replaced.
However, existing LSM-tree based KV-stores have not been able to take advantage of this inherited immutability.
Instead, sorted views are repeatedly reconstructed at search time and immediately discarded afterward,
which leads to poor search performance due to excessive computation and I/O.
The motivation of REMIX is to exploit the immutability of table files by retaining the sorted view
of the underlying tables and reusing them for future searches.

For I/O efficiency,
the LSM-tree based KV-stores employ memory-efficient metadata formats,
including sparse indexes and Bloom filters~\cite{bloom}.
If we record every key and its location to retain the sorted views in a store,
the store's metadata could be significantly inflated, leading to compromised performance for both reads and writes.
To avoid this issue, the REMIX data structure must be space-efficient.

\subsection{The REMIX Data Structure}
\label{sec:remix-ds}

\begin{figure}[t!]
    \centering
    \includegraphics[width=.95\columnwidth]{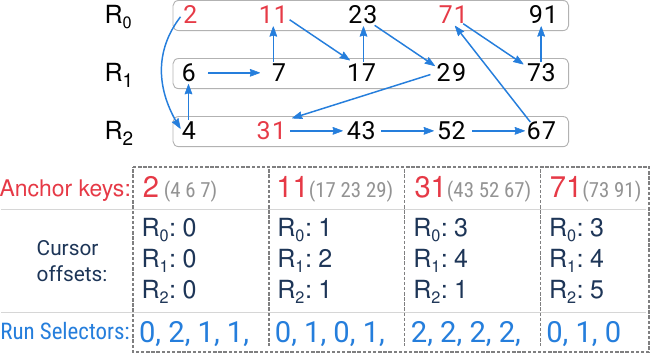}
    \caption{A sorted view of three sorted runs with REMIX}
    \label{fig:design}
\end{figure}

The top of Figure~\ref{fig:design} shows an example of a sorted view containing three sorted runs, $R_0$, $R_1$, and $R_2$.
The sorted view of the three runs is illustrated by the arrows, forming a sequence of 15 keys.
To construct a REMIX, we first divide the keys of a sorted view into segments, each containing a fixed number of keys.
Each segment is attached with an \emph{anchor key}, a set of \emph{cursor offsets}, and a set of \emph{run selectors}.
An anchor key represents the smallest key in the segment.
All the anchor keys collectively form a sparse index on the sorted view.
Each cursor offset corresponds to a run and records the position of the smallest key in the run
that is equal to or greater than the segment's anchor key.
Each key in a segment has a corresponding run selector, which indicates the run where the key resides.
The run selectors encode the sequential access path of the keys on the sorted view,
starting from the anchor key of the segment.

An iterator for a REMIX does not use a min-heap.
Instead, an iterator contains a set of cursors and a \emph{current} pointer.
Each cursor corresponds to a run and points to the location of a key in the run.
The current pointer points to a run selector, which selects a run,
and the cursor of the run determines the key currently being reached.

It takes three steps to seek a key using an iterator on a REMIX.
First, a binary search is performed on the anchor keys to find the \emph{target segment}
whose range covers the seek key, satisfying $anchor\_key \le seek\_key$.
Second, the iterator is initialized to point to the anchor key.
Specifically, the cursors are positioned using the cursor offsets of the segment,
and the current pointer is set to point to the first run selector of the segment.
Finally, the target key can be found by scanning linearly on the sorted view.
To advance the iterator, the cursor of the current key is advanced to skip the key.
Meanwhile, the current pointer is also advanced to point to the next run selector.
After a seek operation, the subsequent keys on the sorted view (within and beyond the target segment)
can be retrieved by advancing the iterator in the same manner.

Here is an example of a seek operation.
As shown in Figure~\ref{fig:design}, the four boxes on the bottom represent the REMIX metadata that encodes the sorted view.
Note that the keys in parentheses are not part of the metadata.
To seek key 17, the second segment, which covers keys $(11, 17, 23, 29)$, is selected with a binary search.
Then the cursors are placed on keys 11, 17, and 31 in $R_0$, $R_1$, and $R_2$, respectively,
according to the segment's cursor offsets ($(1, 2, 1)$).
Meanwhile, the current pointer is set to point to the first run selector of the segment ($0$, the fifth selector in the figure),
indicating that the current key (11) is under the cursor of $R_0$.
Since $11 < 17$, the iterator needs to be advanced to find the smallest key $k$ satisfying $k\ge 17$.
To advance the iterator, the cursor on $R_0$ is first advanced so that it skips key 11 and is now on key 23.
The cursor offsets of the iterator now become 2, 2, and 1.
Then, the current pointer is advanced to the second run selector of the segment ($1$, the sixth selector in the figure).
The advanced iterator selects $R_1$, and the current key 17 under the cursor of $R_1$ is the target key.
This concludes the seek operation.
The subsequent keys (23, 29, 31, \dots) on the sorted view can be retrieved
by repeatedly advancing the iterator.

\subsection{Efficient Search in a Segment}
\label{sec:remix-bsearch}

A seek operation initializes the iterator with a binary search on the anchor keys to find the target segment
and scans forward on the sorted view to look for the target key.
Increasing the segment size can reduce the number of anchor keys and speed up the binary search.
However, it can slow down seek operations because scanning in a large target segment needs to access more keys on average.
To address the potential performance issue, we also use binary search within a target segment to minimize the search cost.

\paragraph{Binary Search}
To perform binary search in a segment, we must be able to randomly access every key in the segment.
A key in a segment belongs to a run, as indicated by the corresponding run selector.
To access a key, we need to place the cursor of the run in the correct position.
This can be done by counting the number of occurrences of the same run selector in the segment
prior to the key and advancing the corresponding cursor the same number of times.
The number of occurrences can be quickly calculated on the fly using SIMD instructions on modern CPUs.
The search range can be quickly reduced with a few random accesses in the segment
until the target key is identified.
To conclude the seek operation,
we initialize all the cursors using the occurrences of each run selector prior to the target key.

\begin{figure}[b!]
    \centering
    \includegraphics[width=\columnwidth]{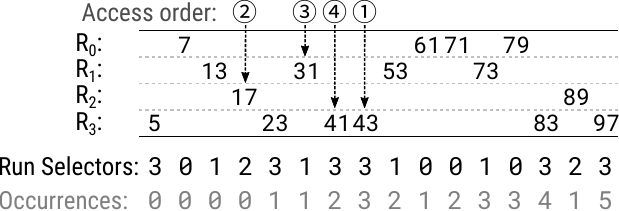}
    \caption{An example of binary search in a segment. The circled numbers indicate the access order of the keys.}
    \label{fig:bsearch}
\end{figure}

Figure~\ref{fig:bsearch} shows an example of a segment having 16 run selectors.
The number shown below each run selector represents the number of occurrences of the same run selector prior to its position.
For example, 41 is the third key in $R_3$ in this segment, so the corresponding number of occurrences is 2 (under the third ``3'').
To access key 41, we initialize the cursor of $R_3$ and advance it twice to skip 5 and 23.

To seek key 41 in the segment in Figure~\ref{fig:bsearch},
keys 43, 17, 31, and 41 will be accessed successively during the binary search,
as shown by the arrows and the circled numbers.
Key 43 is the eighth key in the segment and the fourth key of $R_3$ in the segment.
To access key 43, we initialize the cursor of $R_3$ and advance it three times to skip keys 5, 23, and 41.
Then, key 17 can be accessed by reading the first key on $R_2$ in this segment.
Similarly, 31 and 41 are the second and third keys on $R_1$ and $R_3$, respectively.
In the end, all the cursors of the iterator are initialized to point to the correct keys.
In this example, the cursors will finally be at keys 61, 53, 89, and 41, where 41 is the current key.

\paragraph{I/O Optimization}
Performing binary search in a segment can minimize the number of key comparisons.
However, the keys on the search path may reside in different runs
and must be retrieved with separate I/O requests if the respective data blocks are not cached.
For example, the search in Figure~\ref{fig:bsearch} only needs four key comparisons but has to access three runs.
In fact, it is likely that keys 41, 43, and a few other keys of $R_3$ belong to the same data block.
Accordingly, after a key comparison,
the search can leverage the remaining keys in the same data block
to further reduce the search range before it has to access a different run.
In this way, each of the six keys in $R_3$ can be found without accessing any other runs.
When searching for key 79, for example, accessing $R_3$ can narrow down the search to the range between key 43 and key 83,
where key 79 can be found in $R_0$ after a key comparison with key 71.

\subsection{Search Efficiency}
\label{sec:remix-eff}

REMIXes improve range queries in three aspects.

\paragraph{REMIXes find the target key using one binary search.}
A REMIX provides a sorted view of multiple sorted runs.
Only one binary search on a REMIX is required to position the cursors on the target keys in multiple runs.
Whereas in a traditional LSM-tree based KV-store,
a seek operation requires a number of binary searches on each individual run.
For example, suppose a store with four equally-sized runs has $N$ keys in each run.
A seek operation without a REMIX requires $4\times\log_2 N$ key comparisons,
while it only takes $\log_2{4N}$, or $2+\log_2 N$ key comparisons with a REMIX.

\paragraph{REMIXes move the iterator without key comparisons.}
An iterator on a REMIX directly switches to the next (or the previous) KV-pair by
using the prerecorded run selectors to update the cursors and the current pointer.
This process does not require any key comparisons.
Reading a KV-pair can also be avoided if the iterator skips the key.
In contrast, an iterator in a traditional LSM-tree based KV-store maintains a min-heap
to sort-merge the keys from multiple overlapping sorted runs.
In this scenario, a next operation requires reading keys from multiple runs for comparisons.

\paragraph{REMIXes skip runs that are not on the search path.}
A seek operation with a REMIX requires a binary search in the target segment.
Only those sorted runs containing the keys on the search path will be accessed at search time.
In the best scenario, if a range of target keys reside in one run, such as the segment $(31, 43, 52, 67)$
in Figure~\ref{fig:design}, only one run ($R_2$ in the example) will be accessed.
However, a merging iterator must access every run in a seek operation.

Furthermore, the substantially reduced seek cost allows for
efficient point queries (e.g., GET) on multiple sorted runs indexed by a REMIX without using Bloom filters.
We extensively evaluate the point query efficiency in \S\ref{sec:eval-remix}.

\subsection{REMIX Storage Cost}
\label{sec:remix-cost}

REMIX metadata consists of three components: anchor keys, cursor offsets, and run selectors.
We define $D$ to be the maximum number of keys in a segment.
A REMIX stores one anchor key for every $D$ keys, requiring $1/D$ of the total key size in a level on average.
Assuming the size of a cursor offset is $S$ bytes,
a REMIX requires $S\times H$ bytes to store the cursor offsets for every $D$ keys,
where $H$ denotes the number of runs indexed by a REMIX.
A run selector requires $\lceil\log_2(H)\rceil$ bits.
Adding all the three parts together, a REMIX is expected to store
$\left((\bar{L} + SH)/D + \lceil\log_{2}(H)\rceil/8\right)$ bytes/key,
where $\bar{L}$ is the average anchor key size.

\setlength{\tabcolsep}{2pt}
\begin{table}[b]
\centering
\caption{REMIX storage cost with real-world KV sizes.
BI stands for Block Index. BF stands for Bloom Filter.
The last column shows the size ratio of REMIX to the KV data.}
\label{tab:remix-cost}
\begin{tabular}{l c c|c c|c c c|r}
\hline
\multirow{3}{*}{\makecell{Work-\\load\\{\cite{fb-workload12,fb-workload20}}}} &
\multirow{3}{*}{\makecell{Avg.\\Key\\Size}} &
\multirow{3}{*}{\makecell{Avg.\\Value\\Size}} &
\multicolumn{5}{c|}{Bytes/Key} &
\multirow{3}{*}{\makecell[c]{\large{$\frac{\text{REMIX}}{\text{data}}$}\\{}}} \\
\cline{4-8}
 & & & \multicolumn{2}{c|}{SSTable} & \multicolumn{3}{c|}{REMIX ($H$=8)} & \\
\cline{4-8}
 & & & BI & BI+BF & $D$=16 & 32 & 64 & \makecell[c]{($D$=32)} \\
\hline
UDB     & 27.1  & 126.7 &  1.2 & 2.4 & 4.1 & 2.2 & 1.3 &  1.44\%  \\
Zippy   & 47.9  & 42.9  &  1.2 & 2.4 & 5.4 & 2.9 & 1.6 &  3.16\%  \\
UP2X    & 10.45 & 46.8  &  0.2 & 1.5 & 3.0 & 1.7 & 1.0 &  2.97\%  \\
USR     & 19    & 2     &  0.1 & 1.4 & 3.6 & 2.0 & 1.2 &  9.38\%  \\
APP     & 38    & 245   &  2.9 & 4.2 & 4.8 & 2.6 & 1.5 &  0.91\%  \\
ETC     & 41    & 358   &  4.4 & 5.6 & 4.9 & 2.7 & 1.5 &  0.67\%  \\
VAR     & 35    & 115   &  1.4 & 2.7 & 4.6 & 2.5 & 1.4 &  1.65\%  \\
SYS     & 28    & 396   &  3.3 & 4.6 & 4.1 & 2.3 & 1.3 &  0.53\%  \\
\hline
\end{tabular}
\end{table}

We estimate the storage cost of a REMIX using the average KV sizes
publicly reported in Facebook's production KV workloads~\cite{fb-workload12,fb-workload20}.
In practice, $S$ is implementation-defined, and $H$ depends on the number of tables being indexed.
In the estimation, we use cursor offsets of 4 bytes ($S=4$) so that a cursor offset can address 4\,GB space for each sorted run.
We set the number of sorted runs to 8 ($H=8$).
With these practical configurations, a REMIX stores $\left((\bar{L} + 32)/D + 3/8\right)$ bytes/key.

Table~\ref{tab:remix-cost} shows the REMIX storage costs for each workload with different $D$ ($D=$16, 32, and 64).
For comparison, it also shows the storage cost of the block index (BI) and Bloom filter (BF)
of the SSTable format in LevelDB and RocksDB.
Note that table files indexed by REMIXes do not use block indexes or Bloom filters.
An SSTable stores a key and a block handle for each 4\,KB data block.
The block index storage cost is estimated by dividing
the sum of the average KV size and an approximate block handle size (4\,B)
by the estimated number of KV-pairs in a 4\,KB block.
Bloom filters are estimated as 10 bits/key.
The REMIX storage costs vary from 1.0 to 5.4 bytes/key for different $D$ and $\bar{L}$ values.
For every key size, increasing $D$ can substantially reduce the REMIX storage cost.
The last column $\left(\frac{\text{REMIX}}{\text{data}}\right)$ shows the size ratio of a REMIX to its indexed KV data.
In the worst case (the USR store), the REMIX's size is still less than 10\% of the KV data's size.
\section{RemixDB}
\label{sec:db}

To evaluate the REMIX performance, we implement an LSM-tree based KV-store named RemixDB.
RemixDB employs the tiered compaction strategy to achieve the best write efficiency~\cite{dosto18}.
Real-world workloads often exhibit high spatial locality~\cite{fb-workload12,fb-workload20,twitter-workload20}.
Recent studies have shown that a partitioned store layout
can effectively reduce the compaction cost under real-world workloads~\cite{SLM-DB,evendb20}.
RemixDB adopts this approach by dividing the key space into partitions of non-overlapping key ranges.
The table files in each partition are indexed by a REMIX, providing a sorted view of the partition.
In this way, RemixDB is essentially a single-level LSM-tree using tiered compaction.
RemixDB not only inherits the write efficiency of tiered compaction
but also achieves efficient reads with the help of REMIXes.
The point query operation (GET) of RemixDB performs a seek operation
and returns the key under the iterator if it matches the target key.
RemixDB does not use Bloom filters.

\begin{figure}[b!]
    \centering
    \includegraphics[width=.9\columnwidth]{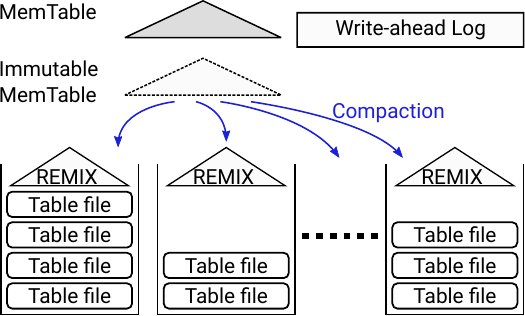}
    \caption{Overview of RemixDB}
    \label{fig:sys}
\end{figure}

Figure~\ref{fig:sys} shows the system components of RemixDB.
Similarly to LevelDB and RocksDB, RemixDB buffers updates in a MemTable.
Meanwhile, the updates are also appended to a write-ahead log (WAL) for persistence.
When the size of the buffered updates reaches a threshold,
the MemTable is converted into an immutable MemTable for compaction,
and a new MemTable is created to receive updates.
A compaction in a partition creates a new version of the partition
that includes a mix of new and old table files and a new REMIX file.
The old version is garbage-collected after the compaction.

In a multi-level LSM-tree design, the size of a MemTable is often only tens of MBs,
close to the default SSTable size.
In a partitioned store layout, larger MemTables can accumulate more updates
before triggering a compaction~\cite{evendb20,flodb17}, which helps to reduce WA.
The MemTables and WAL have near-constant space cost,
which is modest given the large memory and storage capacity in today’s datacenters.
In RemixDB, the maximum MemTable size is set to 4 GB.
In the following, we introduce the file structures (\S\ref{sec:db-files}),
the compaction process (\S\ref{sec:db-comp}), and the cost and trade-offs of using REMIXes (\S\ref{sec:db-rebuild}).

\subsection{The Structures of RemixDB Files}
\label{sec:db-files}

\begin{figure}[h]
    \centering
    \includegraphics[width=\columnwidth]{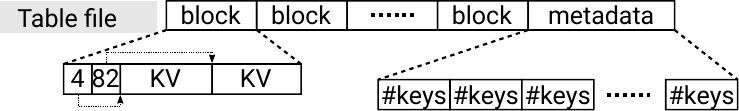}
    \caption{Structure of a table file in RemixDB}
    \label{fig:tablefile}
\end{figure}

\paragraph{Table Files}
Figure~\ref{fig:tablefile} shows the table file format in RemixDB.
A data block is 4\,KB by default.
A large KV-pair that does not fit in a 4\,KB block
exclusively occupies a \emph{jumbo} block that is a multiple of 4\,KB.
Each data block contains a small array of its KV-pairs' block offsets
at the beginning of the block for randomly accessing individual KV-pairs.

The metadata block is an array of 8-bit values, each recording the number of keys in a 4\,KB block.
Accordingly, a block can contain up to 255 KV-pairs.
In a jumbo block, except for the first 4\,KB,
the remaining ones have their corresponding numbers set to 0
so that a non-zero number always corresponds to a block's head.
With the offset arrays and the metadata block,
a search can quickly reach any adjacent block and skip an arbitrary number of keys without accessing the data blocks.
Since the KV-pairs are indexed by a REMIX, table files do not contain indexes or filters.

\begin{figure}[h]
    \centering
    \includegraphics[width=\columnwidth]{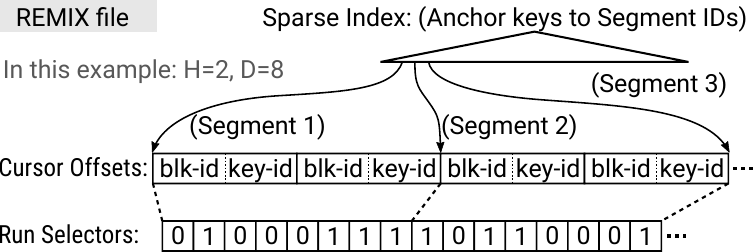}
    \caption{Structure of a REMIX file in RemixDB}
    \label{fig:remixfile}
\end{figure}

\paragraph{REMIX Files}
Figure~\ref{fig:remixfile} shows the REMIX file format in RemixDB.
The anchor keys in a REMIX are organized in an immutable B+-tree-like index
(similar to LevelDB/RocksDB's block index) that facilitates binary searches on the anchor keys.
Each anchor key is associated with a segment ID that identifies the cursor offsets and run selectors of a segment.
A cursor offset consists of a 16-bit block index and an 8-bit key index,
shown as \texttt{blk-id} and \texttt{key-id} in Figure~\ref{fig:remixfile}.
The block index can address up to 65,536 4-KB blocks (256\,MB).
Each block can contain up to 256 KV-pairs with the 8-bit key index.

Multiple versions of a key could exist in different table files of a partition.
A range query operation must skip the old versions and return the newest version of each key.
To this end, in a REMIX,
multiple versions of a key are ordered from the newest to the oldest on the sorted view,
and the highest bit of each run selector is reserved to
distinguish between old and new versions.
A forward scan operation will always encounter the newest version of a key first,
and then the old versions can be skipped by checking the reserved bit of each run selector
without comparing any keys.

If a key has multiple versions, these versions can span two segments.
A search may have to check both the segments to retrieve the newest version of the key.
To simplify searches in this scenario,
we move all the versions of the key forward to the second segment by
inserting special run selectors as placeholders in the first segment when constructing a REMIX.
We also make sure that the maximum number of keys in a segment is equal to or greater than the number of runs indexed by a REMIX ($D \ge H$)
so that every segment is large enough to hold all the versions of a key.

To accommodate the special values mentioned above, each run selector in RemixDB occupies a byte.
The eighth and seventh bits (\texttt{0x80} and \texttt{0x40}) of a run selector
indicate an old version and a deleted key (a tombstone), respectively.
A special value 63 (\texttt{0x3f}) represents a placeholder.
In this way, RemixDB can manage up to 63 sorted runs (0 to 62) in each partition,
which is sufficient in practice.

\subsection{Compaction}
\label{sec:db-comp}

In each partition, the compaction process estimates the compaction cost based on
the size of new data entering the partition and the layout of existing tables.
Based on the estimation, one of the following procedures is executed:
\begin{itemize}
\item Abort: cancel the partition's compaction and keep the new data in the MemTables and the WAL.
\item Minor Compaction: write the new data to one or multiple new tables without rewriting existing tables.
\item Major Compaction: merge the new data with some or all of the existing tables.
\item Split Compaction: merge the new data with all the existing data and split the partition into a few new partitions.
\end{itemize}

\paragraph{Abort}
After a compaction, a partition that sees any new table file will have its REMIX rebuilt.
When a small table file is created in a partition after a minor compaction,
rebuilding the REMIX can lead to high I/O cost.
For example, the USR workload in Table~\ref{tab:remix-cost} has the highest size ratio of REMIX to KV data (9.38\%).
Writing 100\,MB of new data to a partition with 1\,GB of old table files will create a REMIX that is about 100\,MB.
To minimize the I/O cost,
RemixDB can abort a partition's compaction if the estimated I/O cost
is above a threshold.
In this scenario, the new KV data should stay in the MemTables and the WAL until the next compaction.

However, in an extreme case, such as having a workload with a uniform access pattern,
the compaction process cannot effectively move data into the partitions
when most of the partitions have their compactions aborted.
To avoid this problem, we further limit the size of new data that can stay in the MemTables
and WAL to be no more than 15\% of the maximum MemTable size.
The compaction process can abort the compactions that have the highest I/O cost until
the size limit has been reached.

\begin{figure}[t!]
    \centering
    \includegraphics[width=.80\columnwidth]{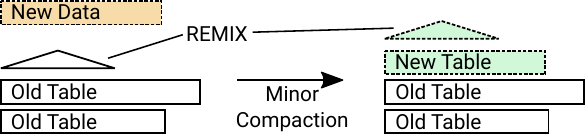}
    \caption{Minor compaction}
    \label{fig:minorcomp}
\end{figure}

\paragraph{Minor Compaction}
A minor compaction writes new KV data from the immutable MemTable into a partition
without rewriting existing table files and rebuilds the REMIX of the partition.
Depending on the new data's size, a minor compaction creates one or a few new table files.
Minor compaction is used when the expected number of table files after the compaction
(number of existing table files plus the estimated number of new table files) is below a threshold $T$,
which is 10 in our implementation.
Figure~\ref{fig:minorcomp} shows a minor compaction example that creates one new table file.

\begin{figure}[t!]
    \centering
    \includegraphics[width=.80\columnwidth]{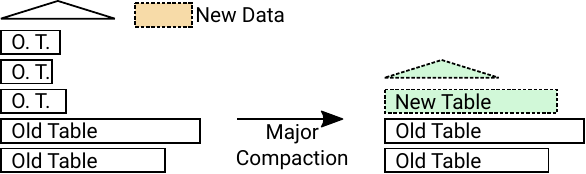}
    \caption{Major compaction}
    \label{fig:majorcomp}
\end{figure}

\paragraph{Major Compaction}
A major (or split) compaction is required when the expected number of table files
in a partition exceeds the threshold $T$.
A major compaction sort-merges existing table files into fewer ones.
With a reduced number of table files, minor compactions can be performed in the future.
The efficiency of a major compaction can be estimated by the ratio of
the number of input table files to the number of output table files.
Figure~\ref{fig:majorcomp} shows a major compaction example.
In this example, the new data is merged with three small table files,
and only one new table file is created after the compaction (ratio=$3/1$).
If the entire partition is sort-merged, the compaction needs to rewrite more data
but still produces three tables (ratio=$5/3$) because of the table file's size limit.
Accordingly, major compaction chooses the number of input files that can produce the highest ratio.

\begin{figure}[t!]
    \centering
    \includegraphics[width=.94\columnwidth]{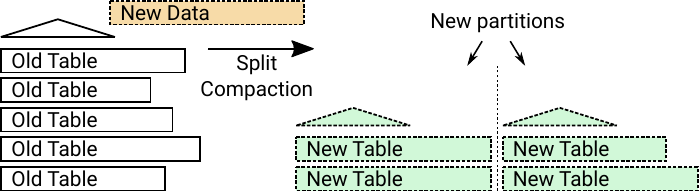}
    \caption{Split compaction}
    \label{fig:splitcomp}
\end{figure}

\paragraph{Split Compaction}

Major compaction may not effectively reduce the number of tables in a partition filled with large tables,
which can be predicted by a low estimated input/output ratio, such as $10/9$.
In this case, the partition should be split into multiple partitions so that
the number of tables in each partition can be substantially reduced.
Split compaction sort-merges new data with all the existing table files in the partition
and produces new table files to form several new partitions.
Figure~\ref{fig:splitcomp} shows a split compaction example.
To avoid creating many small partitions in a split compaction,
the compaction process creates $M$ ($M=2$ by default)
new table files in a partition before switching to the next partition.
In this way, a split compaction creates $\lceil E/M\rceil$ new partitions,
where $E$ is the number of new table files.

\subsection{Rebuilding REMIXes}
\label{sec:db-rebuild}

A partitioned store layout can effectively minimize the compaction cost
under real-world workloads with high spatial locality~\cite{SLM-DB,evendb20}.
Specifically, RemixDB can absorb most of the updates in a few partitions,
and the compactions in the partitions that receive fewer updates can be avoided (See \S\ref{sec:db-comp}).
However, if the workload lacks spatial locality,
it is inevitable that many partitions have to perform compactions with small amounts of updates.
Tiered compaction can minimize writes in these partitions,
but rebuilding the REMIX in a partition still needs to read the existing tables.
In our implementation, RemixDB leverages the existing REMIX in the partition and employs an efficient merging algorithm
to minimize the I/O cost of the rebuilding process.

When rebuilding the REMIX in a partition, the existing tables are already indexed by the REMIX,
and those tables can be viewed as one sorted run.
Accordingly, the rebuilding process is equivalent to sort-merging two sorted runs,
one from the existing data and the other from the new data.
When the existing sorted run is significantly larger than the new one,
the \emph{generalized binary merging algorithm} proposed by Hwang et al.~\cite{taocp3,hwanglin}
requires much fewer key comparisons than sort-merging with a min-heap.
The algorithm estimates the location of each next merge point based on the size ratio between the two sorted runs and search in the neighboring range.
In RemixDB, we approximate the algorithm by using the anchor keys to locate the target segment
containing the merge point and finally applying a binary search in the segment.
In this process, accessing anchor keys does not incur any I/O since they are stored in the REMIX.
A binary search in the target segment reads at most $\log_2{D}$ keys to find the merge point.
All the run selectors and cursor offsets for the existing tables can be derived from the existing REMIX without any I/O.
To create anchor keys for the new segments, we need to access at most one key per segment on the new sorted view.

The read I/O of rebuilding a REMIX is bounded by the size of all the tables in a partition.
The rebuilding process incurs read I/O to the existing tables in exchange for
minimized WA and improved future read performance.
Whether rebuilding a REMIX is cost effective depends on how much write I/O one wants to save
and how much future read performance one wants to improve.
In practice, writes in SSDs are usually slower than reads and can cause permanent damage to the devices~\cite{ssdcontract,optanecontract,endurance10,flash09}.
As a result, reads are more economical than writes, especially for systems having spare I/O bandwidth.
In systems that expect intensive writes with weak spatial locality,
adopting a multi-level tiered compaction strategy~\cite{lsmtrie15,pebbles17}
or delaying rebuilding REMIXes in individual partitions
can reduce the rebuilding cost at the expense of having more levels of sorted views.
Adapting REMIXes with different store layouts is beyond the scope of this paper.
We empirically evaluate the rebuilding cost in RemixDB under different workloads
in \S\ref{sec:eval-remixdb}.

\section{Evaluation}
\label{sec:eval}

In this section, we first evaluate the REMIX performance characteristics (\S\ref{sec:eval-remix}),
and then benchmark RemixDB with a set of micro-benchmarks and 
Yahoo's YCSB benchmark tool that emulates real-world workloads~\cite{ycsb2010} (\S\ref{sec:eval-remixdb}).

The evaluation system runs 64-bit Linux (v5.8.7) on
two Intel Xeon Silver 4210 CPUs and 64\,GB of DRAM.
The experiments run on an Ext4 file system on a 960\,GB Intel 905P Optane PCIe SSD.

\subsection{Performance of REMIX-indexed Tables}
\label{sec:eval-remix}

We first evaluate the REMIX performance.
We implement a micro-benchmark framework that compares the performance of REMIX-indexed tables with SSTables.
The SSTables use Bloom filters to accelerate point queries and employ merging iterators to perform range queries.

\paragraph{Experimental Setup}
In each experiment, we first create a set of $H$ table files ($1\le H\le 16$),
which resemble a partition in a RemixDB or a level in an LSM-tree using tiered compaction.
Each table file contains 64\,MB of KV-pairs, where the key and value sizes are 16\,B and 100\,B, respectively.
When $H\ge 2$, the KV-pairs can be assigned to the tables using two different patterns:
\begin{itemize}
\item \textbf{Weak locality}: each key is assigned to a randomly selected table, which provides weak
access locality since logically consecutive keys often reside in different tables.
\item \textbf{Strong locality}: every 64 logically consecutive keys are assigned to a randomly selected table,
which provides strong access locality since a range query
can retrieve a number of consecutive keys from few tables.
\end{itemize}
Each SSTable contains Bloom filters of 10 bits/key.
A 64\,MB user-space block cache\footnote{LevelDB's LRUCache implementation in \texttt{util/cache.cc}.}
is used for accessing the files.

\begin{figure*}[t!]
 \centering
  \begin{subfigure}{0.32\textwidth}
    \includegraphics[width=\textwidth]{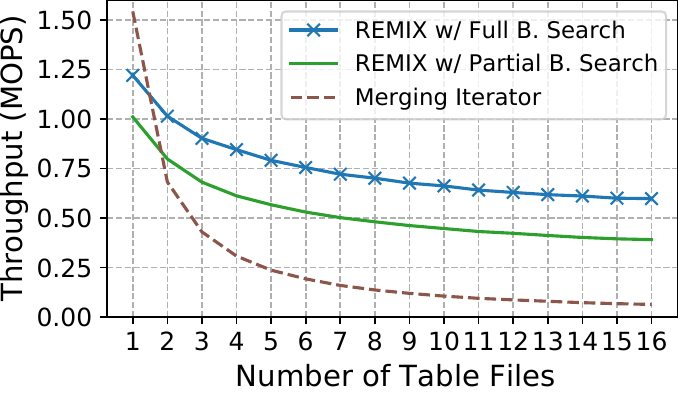}
    \caption{Seek}
    \label{fig:curve_weak_seek}
  \end{subfigure}
  ~
  \begin{subfigure}{0.32\textwidth}
    \includegraphics[width=\textwidth]{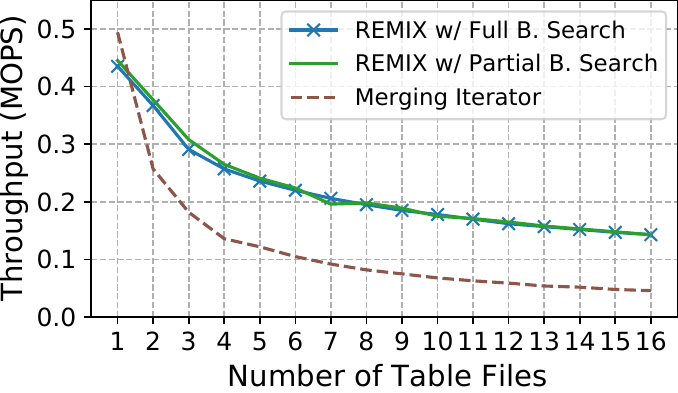}
    \caption{Seek+Next50}
    \label{fig:curve_weak_50}
  \end{subfigure}
  ~
  \begin{subfigure}{0.32\textwidth}
    \includegraphics[width=\textwidth]{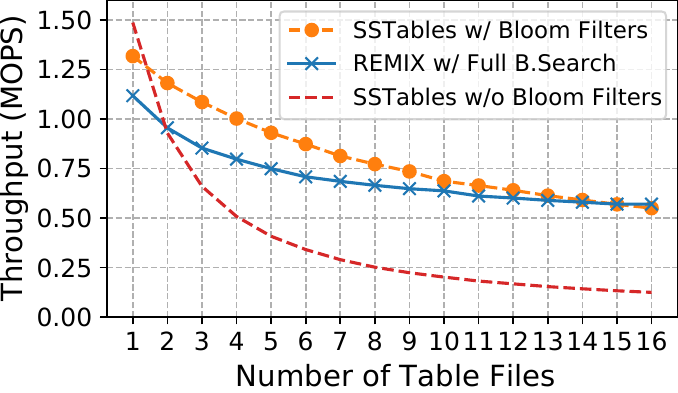}
    \caption{Get}
    \label{fig:curve_weak_get}
  \end{subfigure}
  \caption{Point and range query performance on tables where keys are randomly assigned (weak locality)}
  \label{fig:curve_weak}
\end{figure*}

\begin{figure*}[t!]
 \centering
  \begin{subfigure}{0.32\textwidth}
    \includegraphics[width=\textwidth]{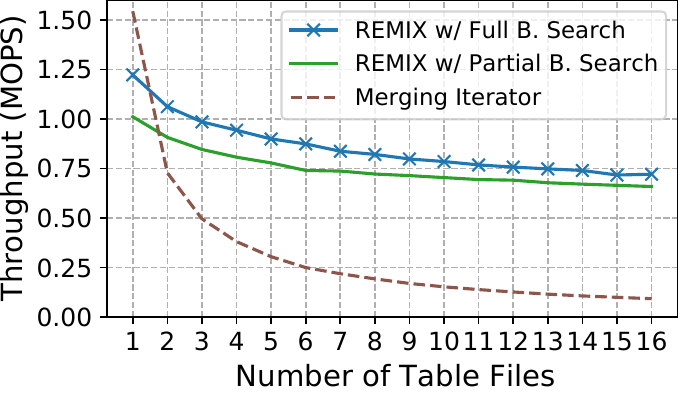}
    \caption{Seek}
    \label{fig:curve_strong_seek}
  \end{subfigure}
  ~
  \begin{subfigure}{0.32\textwidth}
    \includegraphics[width=\textwidth]{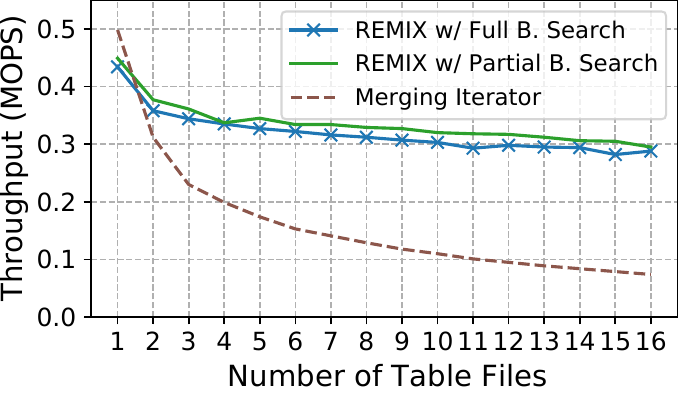}
    \caption{Seek+Next50}
    \label{fig:curve_strong_50}
  \end{subfigure}
  ~
  \begin{subfigure}{0.32\textwidth}
    \includegraphics[width=\textwidth]{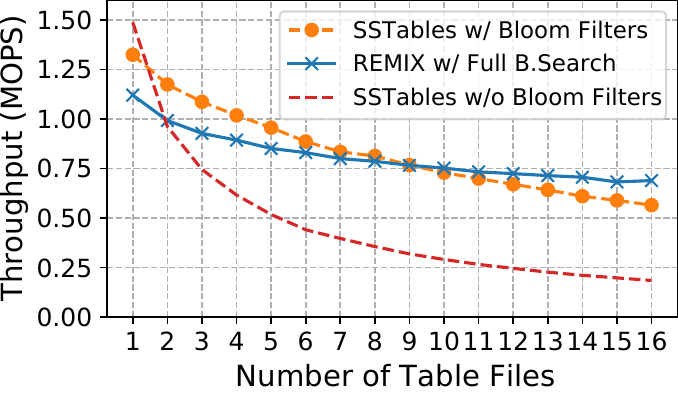}
    \caption{Get}
    \label{fig:curve_strong_get}
  \end{subfigure}
  \caption{Point and range query performance on tables where every 64 keys are assigned to a table (strong locality)}
  \label{fig:curve_strong}
\end{figure*}

We measure the single-threaded throughput of three range and point query operations,
namely Seek, Seek+Next50, and Get,
using different sets of tables created with the above configurations.
A Seek+Next50 operation performs a seek and retrieves the next 50 KV-pairs.
In these experiments, the seek keys are randomly selected following a uniform distribution.
For REMIX, we set the segment size to 32 ($D=32$),
and measure the throughput with its in-segment binary search turned on and off,
denoted by \emph{full} and \emph{partial} binary search, respectively (see \S\ref{sec:remix-bsearch}).
For point queries (Get), we measure the throughput of SSTables with Bloom filters turned on and off.
We run each experiment until the throughput reading is stable.
Figures~\ref{fig:curve_weak} and~\ref{fig:curve_strong} show the throughput results
for tables with weak and strong access locality, respectively.

\paragraph{Seek on Tables of Weak Locality}
Figure~\ref{fig:curve_weak_seek} shows the throughput of seek operations using a REMIX and a merging iterator.
We observe that the throughput with the merging iterator is roughly 20\% higher than
that of a REMIX with full binary search when there is only one table file.
In this scenario, both the mechanisms perform the same number of key comparisons during the binary search.
However, when searching in a segment, the REMIX needs to count the number of occurrences on the fly
and move the iterator from the beginning of the segment to reach a key for comparison,
which is more expensive than a regular iterator.

The throughput of a merging iterator quickly drops as the number of table files increases.
Specifically, the throughput of two tables is 50\% lower than that of one table;
a seek on eight tables is more than $11\times$ slower than a seek on one table.
The seek time of a merging iterator is approximately proportional to the number of table files.
This is because the merging iterator requires a full binary search on every table file.
The REMIX's throughput also decreases with more tables files.
The slowdown is mainly due to the growing dataset that requires
more key comparisons and memory accesses during a search.
However, the REMIX with full binary search achieves increasingly high speedups compared with the merging iterator.
Specifically, The speedups are $5.1\times$ and $9.3\times$ with 8 and 16 table files, respectively.

The REMIX throughput decreases by 20\% to 33\% when the in-segment binary search is turned off (with partial binary search).
In this scenario, a seek has to linearly scan the target segment to find the target key.
With $D=32$, the average number of key comparisons in a target segment is $5$ ($\log_2 D$) with full binary search
and $16$ ($D/2$) with partial binary search.
However, the search cost is still substantially lower than that of a merging iterator.
The REMIX with partial binary search outperforms the merging iterator by $3.5\times$ and $6.1\times$,
with 8 and 16 table files, respectively.

\paragraph{Seek+Next50}
Figure~\ref{fig:curve_weak_50} shows the throughput of range queries that seek and copy
50 KV-pairs to a user-provided buffer.
The overall throughput results are much lower than that in the Seek experiments
because the data copying is expensive.
However, the REMIX still outperforms the merging iterator when there are two or more tables.
The speedup is $1.4\times$, $2.3\times$, and $3.1\times$ with 2, 8, and 16 table files, respectively.
The suboptimal scan performance of the merging iterator is due to the expensive next operation
that requires multiple key comparisons to find the next key on the sorted view.
For each KV-pair copied to the buffer, multiple KV-pairs must be read and compared to find the global minimum.
In contrast, a REMIX does not require any key comparisons in a next operation.

In contrast to the substantial performance gap between the two REMIX curves in Figure~\ref{fig:curve_weak_seek},
the two curves in Figure~\ref{fig:curve_weak_50} are very close to each other.
This phenomenon is the result of two effects:
(1) the next operations dominate the execution time and 
(2) the linear scanning of a seek operation in a segment warms up the block cache, which makes the future next operations faster.

\paragraph{Point Query}
Figure~\ref{fig:curve_weak_get} shows the results of the point query experiments.
The REMIX's curve is slightly lower than its counterpart in Figure~\ref{fig:curve_weak_seek}
because a get operation needs to copy the KV-pair after a seek using the REMIX.
Searching on SSTables with Bloom filters outperforms searching on 
REMIX-indexed table files when there are fewer than 14 tables.
The reasons for the differences are two-fold.
First, a search can be effectively narrowed down to one table file at a small cost of checking the Bloom filters.
Second, searching in an SSTable is faster than on a REMIX managing many more keys.
In the worst case, the REMIX's throughput is 20\% lower than that of Bloom filters (with 3 tables).
Unsurprisingly, the searches with more than two SSTables are much slower without Bloom filters.

\paragraph{Performance with Tables of Strong Locality}
Figure~\ref{fig:curve_strong} shows the range and point query performance on tables with strong access locality.
The results in Figures~\ref{fig:curve_strong_seek} and~\ref{fig:curve_strong_50}
follow a similar trend of their counterparts in Figure~\ref{fig:curve_weak}.
In general, the improved locality allows for faster binary searches since in this scenario
the last few key comparisons can often use keys in the same data block.
However, the throughput of the merging iterator remains low because of the intensive key comparisons that
dominate the search time.
The REMIX with partial binary search improves more than that with full binary search.
This is because improved locality reduces the penalty on the scanning in a target segment,
where fewer cache misses are incurred in each seek operation.

The REMIX point query performance also improves due to the strong locality that speeds up
the underlying seek operations, as shown in Figure~\ref{fig:curve_strong_get}.
Meanwhile, the results of Bloom filters stay unchanged because
the search cost is mainly determined by the false-positive rate and the search cost on individual tables.
As a result, REMIXes are able to outperform Bloom filters when there are more than 9 tables.

\begin{figure}[b!]
  \centering
  \begin{subfigure}{0.48\columnwidth}
    \includegraphics[width=\textwidth]{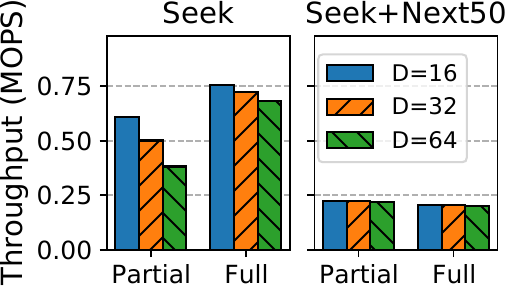}
    \caption{Tables of weak locality}
    \label{fig:remix-table-comp1}
  \end{subfigure}
  ~
  \begin{subfigure}{0.48\columnwidth}
    \includegraphics[width=\textwidth]{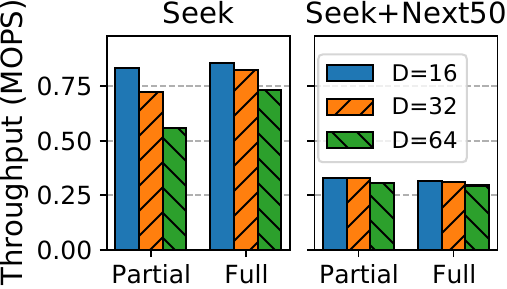}
    \caption{Tables of strong locality}
    \label{fig:remix-table-comp64}
  \end{subfigure}
  \caption{REMIX range query performance with 8 runs}
  \label{fig:remix-table-comp}
\end{figure}

\paragraph{Segment Size ($D$)}
We further evaluate REMIX range query performance using different segment sizes ($D\in\{16,32,64\}$) on eight table files.
The other configuration parameters are the same as in the previous experiments.
Figure~\ref{fig:remix-table-comp} shows the performance results.
The throughput of seek-only operations exhibits the largest variations with different $D$s
when the in-segment binary search is turned off.
This is because the linear scanning in a segment adds a significant cost with a large $D$.
On the other hand, the differences become much smaller with full binary search.
In the meantime, a larger segment size still leads to higher overhead
because of the slower random access speed within a segment.
In the Seek+Next50 experiments, the data copying dominates the execution time and
there are no significant differences when using different $D$s.

\subsection{Performance of RemixDB}
\label{sec:eval-remixdb}

The following evaluates the performance of RemixDB, a REMIX-indexed KV-store based on an LSM-tree.

\paragraph{Experimental Setup}
We compare RemixDB with state-of-the-art LSM-tree based KV-stores,
including Google's LevelDB~\cite{leveldb}, Facebook's RocksDB~\cite{rocksdb}, and PebblesDB~\cite{pebbles17}.
LevelDB and RocksDB adopt the leveled compaction strategy for balanced read and write efficiency.
PebblesDB adopts the tiered compaction strategy with multiple levels for improved write efficiency
at the cost of having more overlapping runs.

LevelDB (\texttt{v1.22}) supports only one compaction thread.
For RocksDB (\texttt{v6.10.2}), we use the configurations suggested in its official Tuning
Guide\footnote{The configuration for \textit{Total ordered database, flash storage}.}~\cite{rocksdb-tuning}.
Specifically,
RocksDB can have at most three MemTables (one more immutable MemTable than LevelDB).
Both RocksDB and RemixDB are configured with four compaction threads.
RemixDB, LevelDB, and RocksDB are all configured to use 64\,MB table files.
For PebblesDB (\texttt{\#703bd01}~\cite{pebblesdbGithub}), we use the default configurations in its \texttt{db\_bench} benchmark program.
For fair comparisons, we disable compression and use a 4\,GB block cache in every KV-store.
All the KV-stores are built with optimizations turned on
(release build).

In our experiments, we choose three value sizes---40, 120, and 400 bytes.
They roughly match the small (ZippyDB, UP2X, USR), medium (UDB, VAR), and large (APP, ETC, SYS)
KV sizes in Facebook's production systems~\cite{fb-workload12,fb-workload20}.
We use 16-byte fixed-length keys, each containing a 64-bit integer using hexadecimal encoding.

\begin{figure}[t!]
    \centering
    \includegraphics[width=\columnwidth]{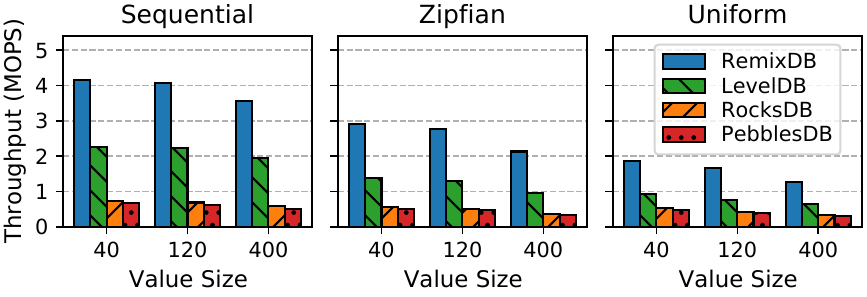}
    \caption{Range query with different value sizes}
    \label{fig:seek-vlen}
\end{figure}

\paragraph{Range Query}
The first set of experiments focuses on how different KV sizes and access patterns
affect the search efficiency of the KV-stores.
In each experiment, we first sequentially load 100 million KV-pairs
into a store using one of the three value sizes.
After loading, we measure the throughput of seek operations using four threads
with three access patterns, namely sequential, Zipfian ($\alpha=0.99$), and uniform.

As shown in Figure~\ref{fig:seek-vlen}, each set of results shows a similar trend.
While RemixDB exhibits the highest throughput,
LevelDB is also at least $2\times$ faster than RocksDB and PebblesDB.
The sequential loading produces non-overlapping table files in every store,
which suggests that a seek operation needs to access only one table file.
However, a merging iterator must check every sorted run in the store
even though they are non-overlapping,
which dominates the execution time of a seek operation if the store has multiple sorted runs.
Specifically, each $L_0$ table in LevelDB and RocksDB is an individual sorted run,
but each $L_i$ ($i>0$) contains only one sorted run;
PebblesDB allows multiple sorted runs in every level.
That being said, LevelDB outperforms RocksDB by at least $2\times$ even though they both use leveled compaction.
We observe that RocksDB keeps several tables (eight in total) at $L_0$ without moving them into a deeper level
during the sequential loading.
In contrast, LevelDB directly pushes a table to a deep level ($L_2$ or $L_3$)
if it does not overlap with other tables,
which leaves LevelDB's $L_0$ always empty.
Consequently, a seek operation in RocksDB needs to sort-merge at least 12 sorted runs on the fly,
while that number is only 3 or 4 in LevelDB.

The seek performance is sensitive to access locality.
A weaker access locality leads to increased CPU and I/O cost on the search path.
In each experiment of a particular value size, the throughput with a uniform access pattern
is about 50\% lower than that of sequential access.
Meanwhile, the performance with sequential access is less sensitive to value size
because the memory copying cost is insignificant.

\begin{figure}[t!]
    \centering
    \includegraphics[width=\columnwidth]{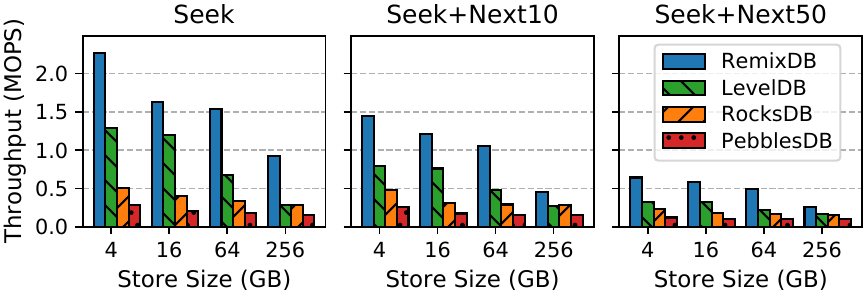}
    \caption{Range query with different store sizes}
    \label{fig:seek-storesize}
\end{figure}

The second set of experiments evaluates the range-scan performance with different store sizes and query lengths.
Each experiment loads a fixed-size KV dataset with 120\,B value size into a store in a random order,
then performs range-scans with four threads using the Zipfian access pattern.
As shown in Figure~\ref{fig:seek-storesize},
RemixDB outperforms the other stores in every experiment.
However, the performance differences among the stores become smaller with longer scans.
The reason is that a long range-scan exhibits sequential access pattern on each sorted run,
where more data have been prefetched during the scan.
In the meantime, the memory-copying adds a constant overhead to every store.

As the store size increases to 256\,GB, the throughput of LevelDB quickly drops to the same level as RocksDB.
Since the stores in the experiments are configured with a 4\,GB block cache,
the cache misses lead to intensive I/Os that dominate the query time.
While RocksDB exhibits high computation cost for having too many $L_0$ tables with a small store size,
the cost is overshadowed by the excessive I/Os in large stores.
Meanwhile, RemixDB maintains the best access locality because it incurs a minimal amount of random accesses
and cache misses by searching on a REMIX-indexed sorted run.

\paragraph{Write}
We first evaluate the write performance of each store by inserting a 256\,GB KV dataset
to an empty store in a random order using one thread.
The dataset has 2 billion KV-pairs, and the value size is 120\,B.
The workload has a uniform access pattern, representing the worst-case scenario of the stores.
We measure the throughput and the total I/O on the SSD.

\begin{figure}[t]
    \centering
    \includegraphics[width=0.73\columnwidth]{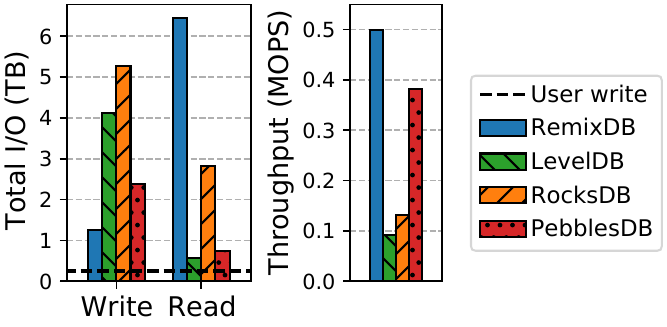}
    \caption{Loading a 256\,GB dataset in random order}
    \label{fig:eval-write-shuffled}
\end{figure}

As shown in Figure~\ref{fig:eval-write-shuffled},
Both RemixDB and PebblesDB show relatively high throughput because they employ
the write-efficient tiered compaction strategy.
Their total write I/O on the SSD are 1.25\,TB and 2.37\,TB,
corresponding to WA ratios of 4.88 and 9.26, respectively.
LevelDB and RocksDB adopt the leveled compaction strategy,
which leads to high WA ratios of 16.1 and 25.6, respectively.
RocksDB and RemixDB have much more read I/O than LevelDB and PebblesDB.
RocksDB employs four compaction threads to exploit the SSD's I/O bandwidth,
resulting in more read I/O than LevelDB due to less efficient block and page cache usage.
LevelDB only supports one compaction thread, and it shows a much lower throughput than RocksDB.
Although RemixDB has more read I/O than RocksDB, the total I/O of RemixDB is less than that of RocksDB.
All told, RemixDB achieves low WA and high write throughput at the cost of increased read I/O.

We further evaluate the write performance of RemixDB under workloads with varying spatial locality.
We use three access patterns, namely sequential, Zipfian ($\alpha=0.99$), and Zipfian-Composite~\cite{evendb20}.
The Zipfian-Composite distribution represents an agglomerate of attributes
in real-world stores~\cite{linkbench13,apache11,fb-workload20}.
With Zipfian-Composite, the prefix of a key (the first 12 bytes)
is drawn from the Zipfian distribution,
and the remainder of the key is drawn uniformly at random.
For each access pattern,
the experiment starts with a 256\,GB store constructed as in the random write experiment
then performs 2 billion updates (with 128\,B values) to the store using the respective access pattern.
We measure the throughput and the total I/O during the update phase.

\begin{figure}[t]
    \centering
    \includegraphics[width=0.82\columnwidth]{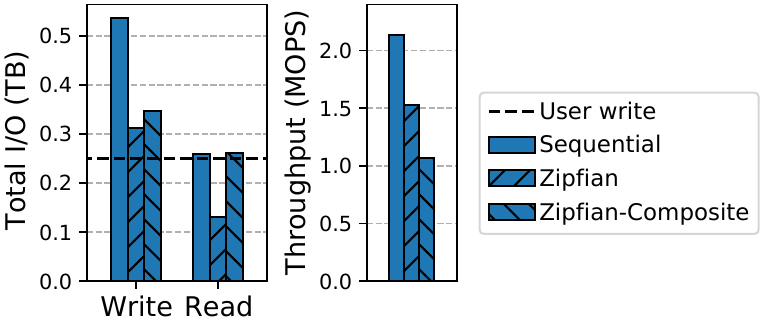}
    \caption{Sequential and skewed write with RemixDB}
    \label{fig:eval-write-skewed}
\end{figure}

As Figure~\ref{fig:eval-write-skewed} shows,
the sequential workload exhibits the highest throughput because 
each round of the compaction only affects a few consecutive partitions in the store.
The write I/O mainly includes logging and creating new table files,
which is about 2$\times$ of the user writes.
The read I/O for rebuilding REMIXes is about the same as the existing data (256\,GB).
Comparatively, with the two skewed workloads,
the repeated overwrites in the MemTable lead to substantially reduced write I/O.
However, the skewed workloads create scattered updates in the key space.
This causes slower updates in the MemTable and more partitions being compacted.
The Zipfian-Composite workload has weaker spatial locality than Zipfian,
resulting in higher compaction I/O cost.

\begin{figure}[t]
    \centering
    \includegraphics[width=\columnwidth]{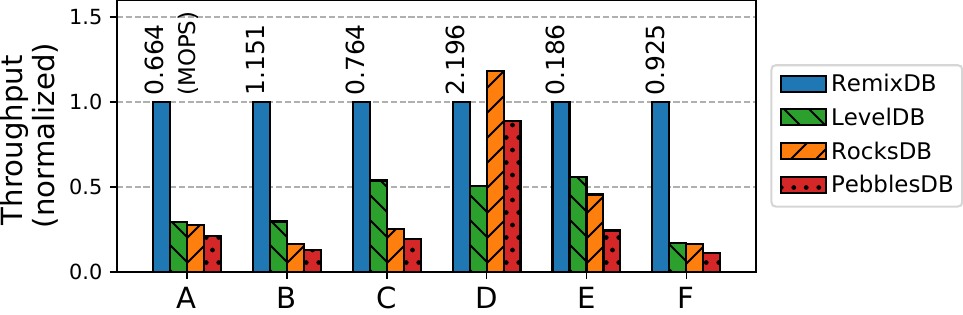}
    \caption{YCSB benchmark results}
    \label{fig:eval-ycsb}
\end{figure}

\paragraph{The YCSB Benchmark}
The Yahoo Cloud Serving Benchmark (YCSB)~\cite{ycsb2010}
is commonly used for evaluating KV-store performance under realistic workloads.
We use the 256\,GB stores constructed in the random-write experiments
and run the YCSB workloads from A to F with four threads.
The details of the workloads are described in Table~\ref{tab:ycsb}.
In workload E, a \textit{Scan} operation performs a seek and retrieves the next 50 KV-pairs.
As shown in Figure~\ref{fig:eval-ycsb},
RemixDB outperforms the other stores except in workload D,
where the read requests (95\%) query the most recent updates produced by
the insertions (5\%).
This access pattern exhibits strong locality,
and most of the requests are directly served from the MemTable(s) in every store.
Meanwhile, LevelDB's performance (1.1\,MOPS) is hindered by slow insertions caused by the single-threaded compaction.

Even though REMIXes do not show an advantage over Bloom filters in the micro-benchmarks
(see Figure~\ref{fig:curve_weak_get}),
RemixDB outperforms the other stores in workloads B and C,
where point query is the dominant operation.
The reason is that a point query in the multi-level LSM-tree
has a high cost selecting candidate tables on the search path.
Specifically, for each $L_0$ table, about two key comparisons are used
to check if the seek key is covered by the table.
If the key is not found at $L_0$, a binary search is used to select a table
at each deeper level $L_i$ ($i \ge 1$) until the key is found.
Furthermore, a Bloom filter's size is about 600\,KB for a 64\,MB table in this setup.
Accessing a Bloom filter performs up to seven random memory accesses,
which leads to excessive cache misses in a large store~\cite{ccfilter14}.
The REMIX-indexed partitions in RemixDB form a globally sorted view,
on which a point query can be quickly answered with a binary search.

\setlength{\tabcolsep}{2pt}
\begin{table}[!t]
\centering
\caption{YCSB Workloads}
\label{tab:ycsb}
\small
\begin{tabular}{l|c|c|c|c|c|c}
\hline
Workload & A & B & C & D & E & F \\
\hline
\multirow{2}*{Operations} &
\multirow{2}{*}{\makecell[l]{ R: 50\% \\ U: 50\% }} & 
\multirow{2}{*}{\makecell[l]{ R: 95\% \\ U: 5\% }} & 
\multirow{2}{*}{\makecell[l]{ R: 100\% }} & 
\multirow{2}{*}{\makecell[l]{ R: 95\% \\ I: 5\% }} & 
\multirow{2}{*}{\makecell[l]{ S: 95\% \\ I: 5\% }} & 
\multirow{2}{*}{\makecell[l]{ R: 50\% \\ M: 50\% }} \\
&&&&&& \\
\hline
Req. Dist. &
\multicolumn{3}{c|}{Zipfian} & 
Latest & 
\multicolumn{2}{c}{Zipfian} 
\\
\hline
\multicolumn{7}{l}{\footnotesize R: Read, U: Update, I: Insert, S:Scan, M: Read-Modify-Write.}
\end{tabular}
\end{table}
\section{Related Work}

\paragraph{Improving Search with Filters}
Bloom filters~\cite{bloom} have been indispensable for LSM-tree based KV-stores
in reducing the computation and I/O costs of point queries on a multi-leveled store
layout~\cite{monkey17}.
However, range queries cannot be handled by Bloom filters
because the search targets are implicitly specified by range boundaries.
Prefix Bloom filters~\cite{prefixFilter03} can accelerate range queries~\cite{rocksdb,leveldb},
but they can only handle closed-range queries on common-prefix keys (with an upper bound).
Succinct Range Filter (SuRF)~\cite{surf18} supports both open-range and closed-range queries.
The effectiveness of using SuRFs is highly dependent on the distribution of keys and query patterns.
Rosetta~\cite{rosetta20} uses multiple layers of Bloom filters to achieve lower
false positive rates than SuRFs.
However, it does not support open-range queries and has prohibitively
high CPU and memory costs with large range queries.
A fundamental limitation of the filtering approach is that
it cannot reduce search cost on tables whose filters produce positive results.
When the keys in the target range are in most of the overlapping tables,
range filters do not speed up queries but cost more CPU cycles in the search path.
In contrast, REMIXes directly attack the problem of having excessive
table accesses and key comparisons when using merging iterators in range queries.
By searching on a globally sorted view, REMIXes improve range query performance
with low computation and I/O cost.

\paragraph{Improving Search with Efficient Indexing}
KV-stores based on B-trees or B+-trees~\cite{lmdb,berkeleydb99}
achieve optimal search efficiency by maintaining a globally sorted view of all the KV data.
These systems require in-place updates on the disk,
which lead to high WA and low write throughput.
KVell~\cite{kvell19} achieves very fast reads and writes by employing
a volatile full index to manage unordered KV data on the disk.
However, the performance benefits come at a cost, including high memory demand and slow recovery.
Similarly, SLM-DB~\cite{SLM-DB} stores a B+-tree~\cite{fastfair18} in non-volatile memory (NVM) to
index KV data on the disk.
This approach does not have the above limitations,
but it requires special hardware support and increased software complexity.
These limitations are also found in NVM-enabled LSM-trees~\cite{novelsm18,matrixkv20}.
Wisckey~\cite{wisckey16} stores long values in a separate log to reduce index size for search efficiency.
However, the approach requires an extra layer of indirection and does not improve performance with small KV-pairs
that are commonly seen in real-world workloads~\cite{fb-workload20,twitter-workload20}.
Bourbon~\cite{bourbon20} trains learned models to accelerate searches on SSTables but does not support string keys.
REMIXes are not subject to these limitations.
They accelerate range queries in write-optimized LSM-tree based KV stores
by creating a space-efficient persistent sorted view of the KV data.

\paragraph{Read and Write Trade-offs}
Dostoevsky and Wacky~\cite{dosto18,wackybush19} navigate LSM-tree based KV-store designs
with different merging policies to achieve the optimal trade-off between reads and writes.
Tiered compaction has been widely adopted for minimizing WA 
in LSM-tree based KV-stores~\cite{pebbles17,scylladb,cassandra}.
Other write-optimized indexes, such as Fractal trees and B$^\epsilon$-trees,
are also employed in KV-store designs~\cite{tokudb,splinterdb20}.
The improvements on write performance often come with mediocre read performance in practice,
especially for range queries~\cite{evendb20}.
REMIXes address the issue of slow reads in tiered compaction.
They achieve fast range query and low WA simultaneously.
\section{Conclusion}
We introduce the REMIX, a compact multi-table index data structure for fast range queries in LSM-trees.
The core idea is to record a globally sorted view of multiple table files for efficient search and scan.
Based on REMIXes, RemixDB effectively improves range query performance while preserving low write amplification using tiered compaction.

\section*{Acknowledgements}

We are grateful to our shepherd William Jannen, the anonymous reviewers,
Xingsheng Zhao, and Chun Zhao, for their valuable feedback.
This work was supported in part by the UIC startup funding and
US National Science Foundation under Grant CCF-1815303. 

\printbibliography

\end{document}